%% file: main.tex
\documentclass[nonacm,sigplan]{acmart}

\usepackage[]{hyperref}
\usepackage{multirow}
\usepackage{booktabs} 
\usepackage{xcolor}
\usepackage{xspace}
\usepackage{formatting/macros}
\usepackage{formatting/results}
\usepackage{balance}
\usepackage{circledsteps}
\usepackage{optidef}
\usepackage[ruled,vlined]{algorithm2e}
\usepackage{float}
\usepackage{natbib}

\begin{document}

\title{Workload-Aware Hardware Accelerator Mining for Distributed Deep Learning Training}

\author{Muhammad Adnan}
\affiliation{%
  \institution{University of British Columbia}
  \city{Vancouver}
  \country{Canada}
}
\email{adnan@ece.ubc.ca}

\author{Amar Phanishayee}
\affiliation{%
  \institution{Microsoft Research}
  \city{Redmond}
  \country{USA}
}
\email{amar@microsoft.com}

\author{Janardhan Kulkarni}
\affiliation{%
  \institution{Microsoft Research}
  \city{Redmond}
  \country{USA}
}
\email{jakul@microsoft.com}

\author{Prashant J. Nair}
\affiliation{%
  \institution{University of British Columbia}
  \city{Vancouver}
  \country{Canada}
}
\email{prashantnair@ece.ubc.ca}

\author{Divya Mahajan}
\affiliation{%
  \institution{Georgia Institute of Technology}
  \city{Atlanta}
  \country{USA}
}
\email{divya.mahajan@gatech.edu}

\begin{abstract}
\input{./body/abstract.tex}
\end{abstract}

\maketitle 
\pagestyle{plain} 

\input{./body/introduction.tex}
\input{./body/background.tex}
\input{./body/challenges.tex}
\input{./body/wham.tex}
\input{./body/distributedsearch.tex}
\input{./body/evaluation.tex}
\input{./body/related_work.tex}
\input{./body/conclusion.tex}

\balance

\bibliographystyle{unsrtnat}
\bibliography{references}

\end{document}

%% file: body/abstract.tex
In this paper, we present a novel technique to search for hardware architectures of accelerators optimized for end-to-end training of deep neural networks (DNNs).
Our approach addresses both single-device and distributed pipeline and tensor model parallel scenarios, latter being addressed for the first time.
The search optimized accelerators for training relevant metrics such as throughput/TDP under a fixed area and power constraints. 
However, with the proliferation of specialized architectures and complex distributed training mechanisms, the design space exploration of hardware accelerators is very large.  
Prior work in this space has tried to tackle this by reducing the search space to either a single accelerator execution that too only for inference, or tuning the architecture for specific layers (e.g., convolution).
Instead, we take a unique heuristic-based critical path based approach to determine the best use of available resources (power and area) either for a set of DNN workloads or each workload individually.
To ensure scalability to for distributed training, we decompose the problem into smaller tasks.
First, we perform local search to determine the architecture for each pipeline and tensor model stage.
Specifically, the system iteratively generates architectural configurations and tunes the design using a novel heuristic-based approach that prioritizes accelerator resources and scheduling to critical operators in a machine learning workload.
Second, to address the complexities of distributed training, the local search selects multiple (k) designs per stage. 
A global search then identifies an accelerator from the top-k sets to optimize training throughput across the stages.
We evaluate this work on 11 different DNN models. 
Compared to a recent inference-only work Spotlight, our method converges to a design in, on average, \convgtimeimprspotlight less time and offers \avgthroughputindividualspotlight higher throughput.
Moreover, designs generated using our method achieve \avgthroughputcommonTPU throughput improvement over TPU architecture.

%% file: body/introduction.tex
\section{Introduction}
\label{sec:introduction}
Special-purpose hardware is well-suited for deep learning models due to their predictable memory accesses and readily parallelizable dataflow patterns~\cite{tpuv4_isca, eyerissv2, brainwave, tabla:hpca, cosmic:micro, dnnweaver:micro}.
As the models become bigger, training them on a single accelerator is infeasible due to their large memory footprint~\cite{gpt, opt, bert}.
This mandates pipeline parallel and/or tensor model parallel training that splits the model across multiple devices~\cite{gpipe, pipedream, pipedream-flush, Alpha21264, megatron}.
However, a general scalable approach to determine the optimal solution for the combined exploration of accelerator architecture and operator execution schedule, in the context of  \emph{distributed training of deep learning models}, is an important yet open problem. 

Prior works in this area have primarily focused on devising solutions for accelerators targeting inference~\cite{fast, prime}.
Some studies scope the architecture search only for matrix multiplication-based operations~\cite{confuciux, hasco, spotlight}.
However, training presents unique challenges compared to inference: the execution graph for training is much larger, it requires greater computational intensity due to the additional backward pass, optimizer, and loss function, and intermediate activations are either stashed or recomputed between forward and backward passes, resulting in a larger memory footprint.
Furthermore, all established pipeline parallel training schemes mandate that backward pass operators be placed and executed on the same device as the forward pass to minimize weight movement across accelerators~\cite{pipedream, gpipe, pipedream-flush, pipemare}. 
As such, the sheer complexity of the search space not only increases due to training, but it also requires co-optimization across forward and backward pass operators and is further compounded by distributed execution.
To tackle the challenge of architecture search in the context of distributed DNN training, we design \wham. 
It answers the following research questions:
\vspace{-2ex}
\begin{enumerate}
    \item \textbf{Individual accelerator design optimization}: What is the optimal architecture, given a specific DNN, under certain area and power constraints while maximizing end-to-end training metrics? Additionally, given a set of DNN workloads, can we identify a common architecture that performs well across all of them? How do algorithmically generated architectures by WHAM compare to previous training accelerators?
    \item \textbf{Global optimization for accelerators in distributed training}: What is the ideal accelerator design for a given set of workloads executing pipeline and/or tensor model parallel training? Are heterogeneous designs obtained by tuning individual accelerators in each stage better than a homogeneous pipeline? 
\end{enumerate}

To address these questions,  \wham leverages the insight that accelerator vendors have converged on offering specialized processors, such as tensor and vector cores, that serve a wide range of common DNN operators~\cite{nvidiaA100, eyeriss, brainwave, brainwave2, tpu}.
Tensor cores execute matrix multiplication-based operations, whereas vector cores execute activation and element-wise operators.
In \wham, each operator in the DNN graph executes on a single computation core.
As a result, \wham employs a tunable architectural template to define the scope of its design space exploration.
%
%
The problem of tuning the hardware architecture boils down to determining the number of tensor and vector cores, their dimensionality, and on-chip buffer sizes.
However, even with a template with only tensor core (maximum size 16 $\times$ 16 and maximum quantity 1), the search space for a MobileNet\_v2 inference accelerator is on the order of $O(10^{72})$~\cite{mobilenet}.
This complexity for training increases to $O(10^{216})$ that includes exploring the tensor core dimensions, L1 register file size, global buffer size, and dataflow of operators~\cite{confuciux}.

Thus, to tackle the scale of accelerator architecture search for distributed training, \wham breaks down the problem into manageable sub-problems.
First, \wham uses existing techniques to partition a model into stages, where each stage is split based on tensor model and/or pipeline parallel, then uses its novel search mechanism to find multiple suitable architectures for each stage in isolation.
Second, to optimize end-to-end training-relevant metrics like throughput or Perf/TDP, \wham does not simply select the best accelerator across each stage.
Instead, it employs the top-k designs for each stage to search for a globally optimized pipeline.

Training is considerably more complex than inference, even for a single accelerator in a stage.  
For every accelerator architecture search, \wham performs a novel heuristic-based search that prioritizes resources and scheduling for throughput-critical operators.
A critical-path analysis offers a bound on the number of tensor and vector core, but this constraint does not affect output quality, as it corresponds to the model's parallelizability limit. 
To avoid iterating through all possible options, \wham strategically trims the search space at every step by eliminating numerous tensor and vector core dimensions based on the feedback from previously explored options.
Overall, with a critical-path-based algorithmic approach and configuration pruning, \wham can reduce the MobileNet\_v2 search space to $O(10^{14})$, a significant reduction compared to black-box approaches.

\begin{figure}
	\centering
    \includegraphics[width=0.45\textwidth]{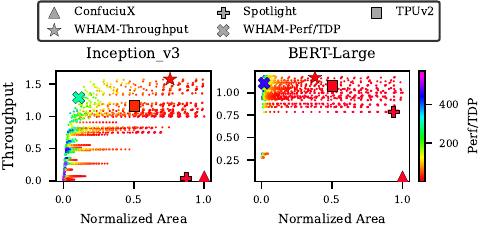}
    \vspace{-1ex}
	\caption{Design space exploration with \wham.}
    \vspace{-3ex}
	\label{fig:dse}
\end{figure}

\wham's design space exploration for Inception\_v3 and Bert-Large, executed on a single accelerator, is illustrated in Figure~\ref{fig:dse}. 
The figure includes a comparison against designs by previous search approaches and a hand-optimized design~\cite{tpuv2}.
With throughput as a metric, \wham converges on a design that maximizes this metric. 
With Perf/TDP as the metric, \wham maximizes Perf/TDP while maintaining a minimum throughput.
Previous work, such as Confuciux and Spotlight, focus solely on inference, thus searches for accelerators based only on the forward pass requirements~\cite{confuciux, spotlight}.
As the figure demonstrates, inference designs are unsuitable for training. The metric of interest varies, the compute requirements for the backward pass differ, and the stashing of intermediate activations is not accounted for.

We evaluate \wham using ten models spanning image classification, translation, and language models. 
For individual accelerator search, on average, the generated designs per workload, provide \avgthroughputindividualconfuciux and \avgthroughputindividualspotlight higher training throughput against ConfuciuX and Spotlight suggested designs while taking \convgtimeimprconfuciux and \convgtimeimprspotlight less time to converge, respectively.
When optimizing an accelerator for a set of DNNs, \wham's common design yields \avgthroughputcommonnvdla and \avgthroughputcommonTPU  better throughput than hand-optimized designs like NVDLA and TPUv2, respectively. 
\wham's top-k based global architecture for distributed training with a pipeline depth of 32, the resulting design optimized per model offers \avgdistthroughputindividual higher throughput and \avgdistperftdpindividual better Perf/TDP compared to the TPUv2 accelerator.

%% file: body/background.tex
\section{Background}

\wham addresses an important yet open question: How to optimize the accelerator architecture and their corresponding runtime operator schedules for distributed training of large deep learning models? 
This section provides an overview of training, focusing on the challenges it poses for architecture search. 
We also outline several established strategies used to facilitate distributed and memory-efficient execution, which, in turn, complicates the search process further.

\subsection{Architecture Search for Training vs Inference}

The training graph comprises three broad passes - forward, backward, and parameter update.
Unlike inference, training stashes activations in the \emph{forward pass} for consumption during the \emph{backward pass}.
Intermediate activations are stored in the memory and fetched during the backward pass for loss gradient calculation.
Thus, intermediate activations have a shorter lifespan in inference, lasting until the next consuming operation. 
In contrast, they persist for a relatively longer duration in training until the execution of the corresponding operator's backward pass.

Previous research in architecture search has primarily focused on inference-only accelerators, which do not account for training passes or consider the memory needed for storing activations~\cite{fast, prime}. 
Other studies have mainly optimized within matrix multiplication-based Convolution and GEMM operations, neglecting other operations~\cite{confuciux, timeloop, maestro, interstellar_dataflow_overrated}.
However, pointwise vector and non-linear operations cannot be ignored when evaluating end-to-end training, as they can contribute significantly to the total training time.
For example, Softmax execution time in BERT~\cite{bert}scales quadratically with increasing sequence length and can take up to 30\% of total training time on a TPU architecture~\cite{tpu}.
Moreover, previous approaches to architecture search are computationally expensive as they rely on black-box optimizations~\cite{vizier}, Bayesian optimization~\cite{hasco, spotlight}, evolutionary~\cite{apollo}, model-based learning (MBO)~\cite{mbo} or machine learning based techniques~\cite{prime}, as shown in Table~\ref{tab:related_work}.
These methods search through and evaluate a large number of configurations to select the best option, albeit only for inference.
\wham instead accounts for the entire training graph, both tensor and vector operations, co-locates forward, corresponding backward and optimizer nodes on the same accelerator, and optimizes across them.

\input{./body/tables/related_work.tex}

\subsection{Distributed Training}

For training, different degrees of parallelism can be employed.
Data parallel training is a common strategy where model replicas are executed on multiple devices~\cite{dataparallel}. 
While effective in accelerating the training of many models, the recent trend of growing model sizes necessitates splitting the model across devices.
This requires other parallelization schemes such as pipeline- and model-parallel training.
Pipeline parallel training is an established method for dividing a model across a pipeline of stages, alleviating memory capacity requirements while maintaining training fidelity. 
Various pipelining strategies, such as Pipedream~\cite{pipedream-flush, pipedream}, GPipe~\cite{gpipe}, and Pipemare~\cite{pipemare} exhibit different memory footprints as the order of micro-batches and when the pipeline is flushed, varies.
Another technique to further split the model to reduce memory requirements, and support parallel execution, is called model parallelism, specifically tensor model parallelism, where weights of a single operation are split across devices~\cite{megatron, flexflow}.
Training large models involves combining some or all of these parallelization techniques.
Pipeline and model parallel training execute different parts of the model across devices, thus is the focus of this work.
\emph{As such, \wham is the first work to support architectural exploration for pipeline parallel training through a combined architectural optimization across pipeline stages.}

\subsection{Device placement for Distributed Execution}
Device placement and determining the distribution strategy is not the focus of this work, thus \wham leverages existing techniques to split a model in pipeline and model parallel fashion.
For pipeline parallel, while \wham can support complex techniques, such as reinforcement learning~\cite{deviceplacement-rl}, dynamic programming~\cite{pip, piper}, and randomized ~\cite{flexflow} search, in this work, we evaluate \wham using a memory-capacity-based model partitioning scheme.
For model parallel, which is commonly employed for large language models, \wham considers the well established Megatron strategy~\cite{megatron}.
Megatron style model parallel splits the attention layer to reduce the memory footprint per device.
Overall, \wham focuses on conducting an accelerator architecture search independent of the pipeline or tensor model scheme and device placement strategies.
Pipeline and tensor model parallel scheme and device placement strategy are inputs to the search.

%% file: body/tables/related_work.tex
\begin{table}

\begin{center}
\caption{Scope of prior works and \wham}
\resizebox{1\columnwidth}{!}{
\begin{tabular}{|l|c|c|}
\hline
\textbf{Works}&\textbf{Search Support and Operators}&\textbf{Search Technique} \\
\hline
FAST~\cite{fast} & Inference (Tensor + Vector Operators) & Black-box (Vizier) \\
ConfuciuX~\cite{confuciux} & Inference (Tensor Operators) & RL and Genetic Algorithm\\
HASCO~\cite{hasco} & Inference (Tensor Operators) & Multi-objective Bayesian Optimization \\
PRIME~\cite{prime} & Inference (Tensor + Vector Operators) & Machine Learning-based Surrogate \\
Spotlight~\cite{spotlight} & Inference (Tensor Operators) & Domian Aware Bayesian Optimization \\
\textbf{WHAM} & \textbf{Training/Inference} \textbf{(Tensor + Vector)} & \textbf{Critical-path-based Heuristics and ILP}\\ 
\hline
\end{tabular}}
\label{tab:related_work}
\vspace{-3ex}
\end{center}
\end{table}

%% file: body/challenges.tex
\section{Architectural Search Parameters}

An architectural template defines the realm of \wham's architecture search.
This architectural template covers a wide variety of machine learning accelerators from the literature~\cite{tpu, nvdla, brainwave, dnnweaver:micro, dadiannao:micro:2014, eyeriss, eyerissv2, tabla:hpca, simba}.
This template is based on fundamental units commonly deployed for machine learning execution, tensor core and vector core. 
Tensor cores are 2-D arrays of Processing Engines (PEs), while vector cores consist of 1-D arrays.
Each PE carries out a \emph{scalar operation}, and together as a core, they can execute larger operations, such as convolutions, dot products etc. 
Each core in the computational unit also has dedicated on-chip storage.

\subsection{Architectural Template}

\begin{figure}
	\centering
	\includegraphics[width=0.45\textwidth]{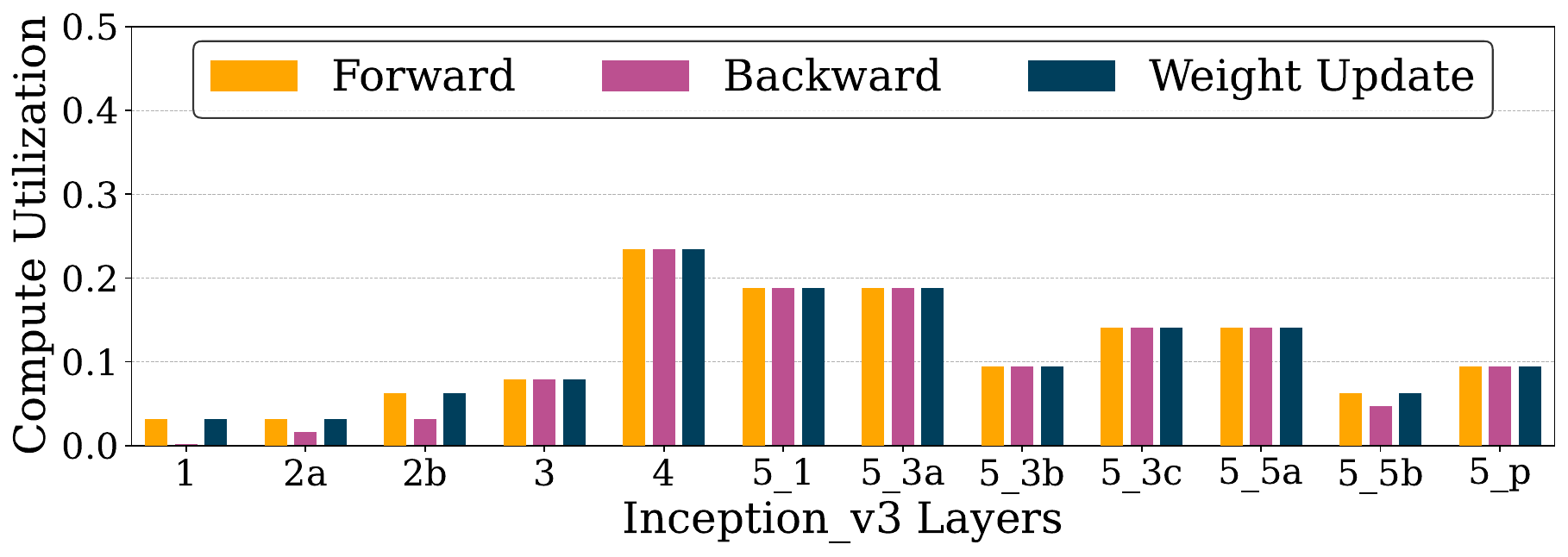}
	\caption{The per-layer utilization of tensor cores and vector cores in Inception\_v3 model using a single tensor (256 $\times$ 256) and vector core (256 wide). The y-axis is capped at 50\%. Layers with fewer channels have lower utilization.}
    \vspace{-2ex}
	\label{fig:motivation}
\end{figure}

To define \wham's architectural template, we look at the evolution of a well-established deep learning accelerator, TPU~\cite{tpu}. TPUv1 comprises a single $256 \times 256$ systolic array in a chip with separate storage for activations and partial sums. TPUv2~\cite {tpuv2} is a training accelerator with a reduced systolic array size of $128 \times 128$, and TPUv3~\cite{tpuv3} is a dual-core chip with each core having two $128 \times 128$ systolic arrays for training.
Although large systolic arrays provide more compute per byte of High-Bandwidth Memory (HBM) bandwidth, they can be inefficient, as numerous workloads fail to fully utilize the $256 \times 256$ systolic array, as demonstrated by Figure~\ref{fig:motivation} for Inception\_v3. 
This observation leads to the fundamental question that \wham explores: what are the appropriate number of cores, size of each core, and number of cores per computational unit for a single or set of machine learning workloads?

\begin{figure}
	\centering
	\includegraphics[width=0.9\columnwidth]     
        {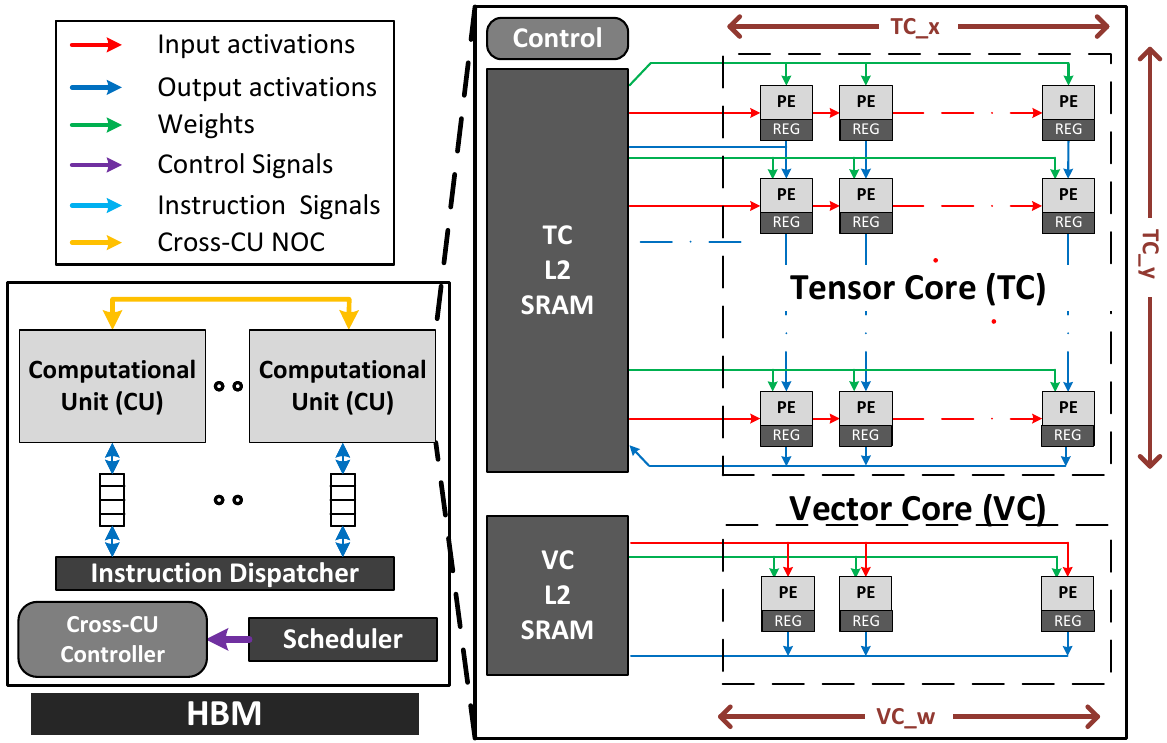}
	\caption{Architectural template. \wham explores TC dimensions, VC width, and number of TCs and VCs.}
    \vspace{-4ex}
	\label{fig:arch_template}
\end{figure}

\wham's architectural template encapsulates these search parameters and builds on prior work~\cite{eyeriss, tpu, brainwave, procrustes, simba, nvdla}.
Figure~\ref{fig:arch_template} illustrates the micro-architecture of the template. 
It consists of computational units, each containing at most one Tensor Core (TC), one Vector Core (VC), or both, performing dense computations. 
Each core's inputs, outputs, and activations are stored in the L2 SRAM.
Activations are stashed for the backward pass in the HBM.
The scheduler generates control signals to execute each operator through the instruction dispatcher and FIFOs. A network-on-chip manages data transfer between cores.
Table~\ref{tab:design_space} displays the tunable parameters of this template, which accommodate a wide range of architectures. 
Each architecture design point is represented as: \emph{$<$\#TC, TC-Dim, \#VC, VC-Width$>$}, indicating the number of TCs, 2-dimensional size of the TC, number of VCs, and 1-dimensional size of the VC, respectively. 
This flexibility enables \wham to explore designs based on the model's compute, memory, and dataflow requirements without being limited to a specific family of accelerators.

\input{body/tables/design_space}

Despite using the template, the search space for training workloads is vast, determined by core dimensions, the number of cores, operator schedules on the accelerator, and exploration of hardware dataflow of individual operators on a core. 
For example, for a moderately sized model like BERT-Base on a single accelerator, the exhaustive search space is on the order of $\sim 10^{2300}$, which is prohibitively expensive.
This search complexity is further complicated by distributed training.
To address this, \wham refines the problem by iteratively searching for core dimensions and quantity. 
It employs an intelligent pruner with a performance feedback loop for the core dimensions to minimize the exploration. 
For core quantity, critical path analysis determines the count based on the parallizability and density of the training graph, assuming infinite resources and no area or power constraints. 
\wham allocates accelerator resources and scheduling to critical operators, significantly impacting the end-to-end metric.
These \wham techniques are described in detail in the next section.

%% file: body/tables/design_space.tex
\begin{scriptsize}
\begin{table}[!ht]
\centering
\caption{Architecture Configuration Parameters}
\resizebox{0.8\columnwidth}{!}{
\begin{tabular}{l c l}
 \hline
 \textbf{Parameter Description} & \textbf{Notation} & \textbf{Range of Values} \\
 \hline
 No. of Tensor Cores & \#TC & 1 to 256 \\
 Tensor Core x dim & TC\_x & 4 to 256 \\
 Tensor Core y dim & TC\_y & 4 to 256 \\
 No. of Vector Cores & \#VC & 1 to 256 \\
 Vector Core width & VC\_w & 4 to 256 \\
 \hline
\end{tabular}}
\label{tab:design_space}
\end{table}
\end{scriptsize}


%% file: body/wham.tex
\section{Accelerator Search with \wham}
\label{sec:local}

\begin{figure}
	\centering
	\includegraphics[width=0.48\textwidth]{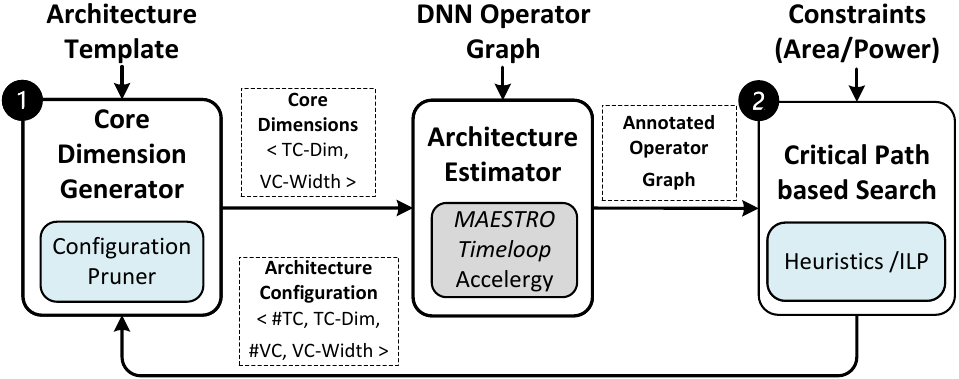}
	\caption{\wham's accelerator search takes an algorithmic approach to prune the large search space for training.}
	\label{fig:local}
    \vspace{-3ex}
\end{figure}

The search process for an accelerator in \wham takes an architectural template and a training operator graph as input, as illustrated in Figure~\ref{fig:local}. 
In deep learning, a model is defined by a pre-set layout that comprises the number of layers, the types of layers (e.g., convolutional, recurrent, dense, transformer), and the connections between them. 
The training operator graph further breaks down these layers into individual dense computations occurring during the forward pass, back-propagated backward pass, parameter update, and loss function.
Each computation in the operator graph is executed on a specific type of hardware core, such as a tensor core, a vector core, or occasionally both. 
In some instances, operators like GEMM (executed on tensor core) and RELU (executed on vector core) are fused to minimize data transfer between off-chip HBM and on-chip memory~\cite{fast, tvm, polymath, yingyang}. 
These fused operators execute on tensor and vector cores simultaneously, enhancing efficiency and reducing latency.

The \wham accelerator search tunes the architectural template for the operator graph it hosts.
This is either a model partition for distributed training or the entire model for single accelerator execution.
The search determines core dimensions and their quantity (represented as \emph{<\#TC, TC-Dim, \#VC, VC-Width>}), and establishes an effective operator schedule that optimizes end-to-end training metrics. 
This schedule efficiently utilizes on-chip resources and a tightly coupled HBM.

To perform the search, the module \circled{1} generates dimensions (\emph{$<$TC-Dim, VC-Width$>$}) starting with the largest configuration that fits within the area constraint.
For each dimension, the operator graph is annotated with the necessary latencies used by the subsequent critical-path-based search.
Then, module \circled{2} determines the number of cores (\emph{$<$\#TC, \#VC$>$}) best suited for the model by leveraging critical operators in the training graph and prioritizing cores and scheduling for those.
This search takes advantage of the insight that the backward pass in training mirrors the dataflow of the forward pass, although with different operators. 
By resolving resource conflicts in the forward pass, \wham's search can potentially mitigate conflicts in the backward pass.

To prune the search, \wham does not explore all possible core dimensions and select the best. Instead, the configuration pruner uses performance feedback from previously explored dimensions to determine early stopping.

\subsection{Core Dimension Generator}
The Dimension Generator iteratively generates \emph{$<$TC-Dim, VC-Width$>$} for Tensor Core Dimension and Vector Core width, respectively. 
Starting with the largest architecture, dimensions are decreased per iteration until convergence. 
Following prior work, we use \emph{<256 $\times$ 256, 256>} as the largest design and explore dimensions in powers of 2 to accommodate common tensor shapes  (batch size, sequence length, hidden size, embedding width) in DNN models. However, \wham supports any step size.
This module utilizes feedback from previous searches to determine whether smaller configurations need evaluation or if the search has concluded, avoiding evaluation across all configurations. 
The configuration pruner is detailed in Section~\ref{sec:pruner} and Figure~\ref{fig:pruning}.

\subsection{Architecture Estimator}

For each \emph{$<$TC-Dim, VC-Width$>$}, the Architecture Estimator annotates the operator graph with essential information from the critical-path-based search. Since each operator executes on one or both cores, only TC-Dim and VC-Width are needed to determine this information.
Each operator in the graph, across forward and backward passes, is annotated with the core type it executes on, latency for execution on this core, and energy expended. 
The latency of each operator allows the rest of the flow to identify latency-critical operators and assess if adding more cores of the required type would resolve resource conflicts during execution.

We use established open-source tools like Timeloop~\cite{timeloop} and MAESTRO~\cite{maestro} to determine the latency of tensor core operators.
For other operators, such as vector and point-wise operations, performance is modeled using a custom model similar to prior work~\cite{fast}. 
The operator latency accounts for compute and data movements from HBM and on-chip memories.
On-chip memory, denoted by \emph{$<$$TC_{L2-SRAM}$, $TC_{L1-REG}$$>$}, is determined based on TC-Dim and the dataflow employed by Timeloop/MAESTRO mapping. 
The \emph{$<$$VC_{L2-SRAM}$$>$} is based on the VC-Width to ensure full vector core utilization. To avoid stalls in the vector core, L2-SRAM is set according to VC-Width.
For energy estimations of each operator, we leverage the established Accelergy~\cite{accelergy} tool.

The Architecture Estimator feeds the annotated operator graph with latencies and energy expended across forward and backward operators to the critical-path-based search.

\subsection{Critical Path-Based Search}
\label{sec:has}
\wham is the first work to propose a critical-path-based approach for architecture search. 
This approach is an alternative to black-box optimizers or reinforcement learning.
The idea of using such critical path-based heuristics is that training graphs can be large and require optimization across co-located forward and backward operators.  
This search technique leverages the insight that auto-grad in training mirrors the forward pass dataflow to the backward pass, where the backward operators correspond to partial derivatives of forward operators.
Based on this insight, the critical path analyzer identifies latency-critical operators and checks for resource conflicts. 
If conflicts are observed, the critical-path-based heuristic adds the required core for that operator, potentially resolving the conflict in both forward and backward passes.
The scheduler prioritizes critical operators, as delaying them would reduce the training throughput.
As a first step, this module determines the theoretically best possible latency and critical operators for each architectural configuration, followed by the search algorithms.

\begin{figure}
	\centering
	\includegraphics[width=0.35\textwidth]{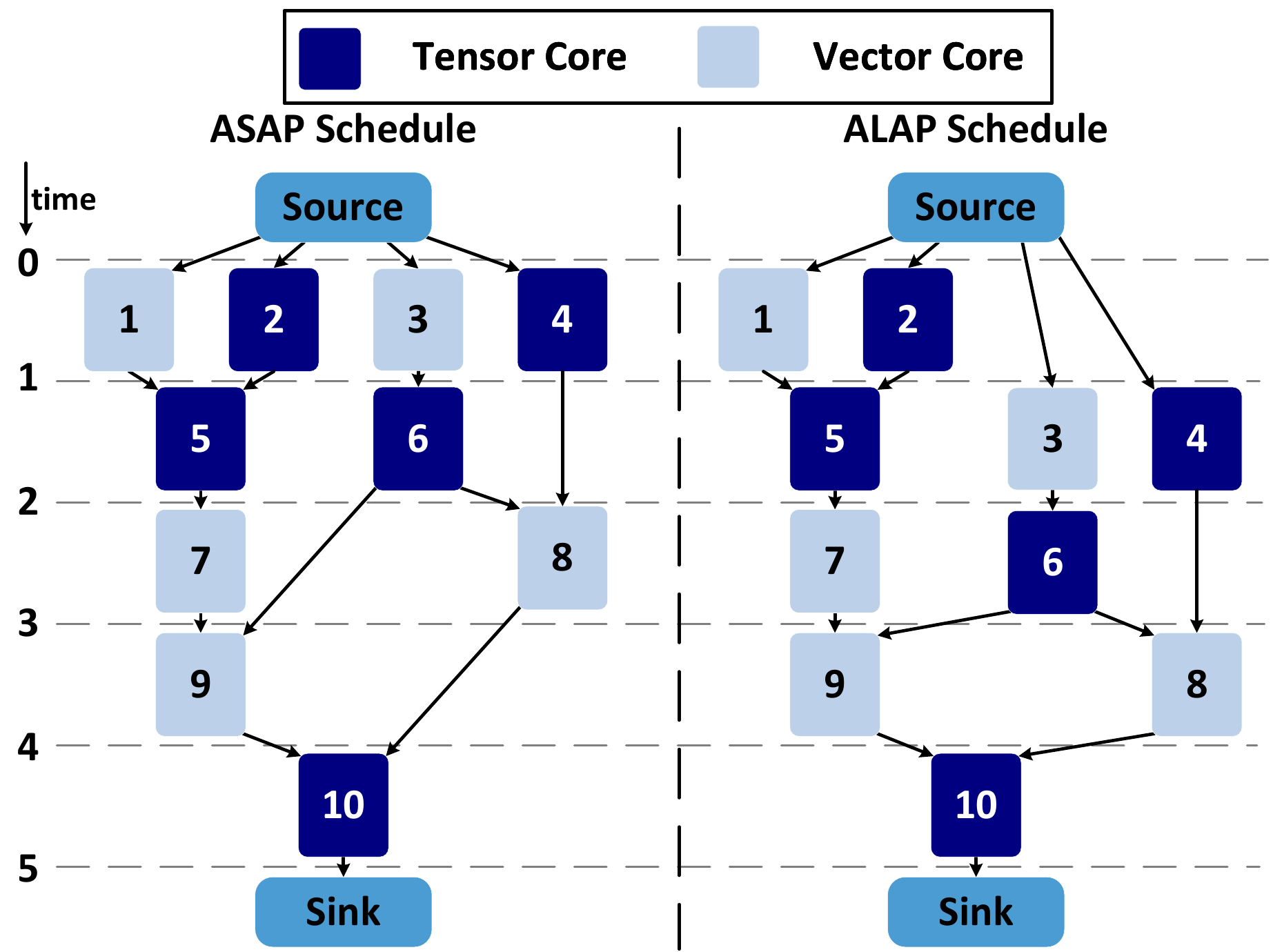}
	\caption{ASAP and ALAP schedules. For simplicity, each operator executes with a unit cycle.}
	\label{fig:critical_path}
    \vspace{-3ex}
\end{figure}

\niparagraph{Theoretical Best Latency and critical operators.}
For every \emph{$<$TC-Dim, VC-Width$>$}, \wham uses operator estimates to determine the theoretically best possible latency a graph can achieve. 
As Soon As Possible (ASAP) scheduling provides the best latency for the operator graph's forward and backward passes. 
As Late as Possible (ALAP) scheduling is also required to determine the critical path. Both ASAP and ALAP schedules presume an infinite number of each core type, as illustrated in Figure~\ref{fig:critical_path}.
ASAP scheduling fully exploits parallelization within the graph by scheduling operators as soon as their predecessors are complete. In contrast, ALAP schedules each operator as late as possible without impacting the overall best latency. 
Operators with the same ASAP and ALAP time are the most critical operators.
These operators do not have any slack in their scheduling window.

\input{./body/heuristics}

\input{./body/ilp}

\subsection{Architecture Configuration Pruner} 
\label{sec:pruner}

\begin{figure}
	\centering
	\includegraphics[width=0.45\textwidth]{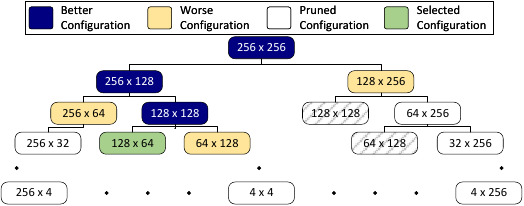}
	\caption{Architectural configuration pruning, each evaluated core dimension executes the heuristics/ILP to determine the \# of cores.}
	\label{fig:pruning}
\end{figure}

For each configuration the dimension generator generates, heuristics or ILP are executed to search for the number of cores. 
Naively exploring every configuration is time-consuming, so \wham employs a novel pruner to reduce the number of core dimensions explored. 
The design space is represented as a binary tree, with the largest dimension at the top level and the next level nodes representing dimensions reduced by the step size. 
The pruner runs for each core type while keeping the other core's configuration constant.
Figure~\ref{fig:pruning} shows this tree design space with top-level design point as \emph{$<$ 256 $\times$ 256 $>$} and step-size of power of $2$ for the tensor core, and fixed vector core width.
The pruner uses a breadth-first algorithmic technique that prunes an entire subtree if one child configuration is better than the parent and the other is worse. 
To avoid selecting a local minimum, a hysteresis level is applied only when all direct child core dimensions perform worse than the parent node. 
In this case, the children are evaluated for multiple sub-levels, and if all these dimensions are worse than the original parent, the entire subtree is pruned. 
This detailed pruning algorithm is shown in Algorithm~\ref{alg:config_gen}. 

\input{./body/algorithms/config_gen}

This technique is based on the insight that if a smaller core dimension does not offer a better training metric than the parent node, there is either insufficient parallelism in the model's operator graph to exploit or the model tensor shapes are not aligned with the architecture configuration. 
In either case, smaller configurations are not beneficial and are pruned. The core dimension generator stops generating new configurations when it reaches leaf configurations or the entire subtree is pruned. Ultimately, \wham's search selects the best architecture based on the training metric from all explored configurations.

\niparagraph{Search Space Comparisons.}
\wham efficiently explores a pruned set of core dimensions, searching for the number of cores and scheduling space for a given operator graph.
Existing toolchains are used for per-operator dataflow exploration~\cite{timeloop, maestro}.
Table~\ref{tab:search_space} compares the search space explored with and without \wham's pruner for both heuristics and ILP, excluding per-operator dataflow search complexity.
The table also compares against unconstrained exhaustive search, where neither the pruner nor the critical path-based algorithmic approach is employed.  

\input{./body/tables/search_space}

\wham significantly reduces the search space compared to the exhaustive due to the critical path-based bounds, even when using optimality-guaranteeing techniques like ILP. 
The pruner eliminates dimensions that cannot offer higher benefits due to graph properties, resulting in the same architecture for both pruned and unpruned searches. 
Across various models, the pruner reduces the search space by order of 10 compared to unpruned searches, decreasing \wham's convergence time by 65\% and 70\% compared to unpruned heuristics and ILP, respectively. 
Convergence time and quality of results are discussed in detail in Section~\ref{sec:evaluation}.

\subsection{\wham-Common}
\wham search conducts architecture optimization for each DNN. 
When optimizing for a set of workloads, the pruner tracks a weighted average of the metric of interest. 
The allows for homogeneity, targeting ASICs based on common compute, data flow, and memory requirements of the workloads. In our evaluation, equal weight is assigned to each workload.

%% file: body/heuristics.tex
\niparagraph{Critical-path guided Heuristics.}
For each core dimension, \wham employs heuristics for tuning the number of cores.
It takes as input the ASAP and ALAP schedules for each forward and backward operator, critical operator information, and a bound on the maximum number of cores required by the training graph.
 These heuristics, called Mirror Conflict Resolution (MCR), start with a single core of \emph{$<$TC-Dim, VC-Width$>$} and iteratively determine which core needs to be added.
Each iteration adds either one tensor or vector core or an entirely new computational unit with both cores.
The criterion for adding a core is: operators are scheduled using a greedy scheduler described below on the current number of TCs and VCs; if a resource conflict causes a delay for an operator beyond its slack in the ALAP schedule, the core that executes the operator is added. 
Fused operators are executed on a computational unit with both tensor and vector cores.
If adding the core/unit to the first conflict does not violate area and power constraints, the change is finalized.
The iterative process of MCR builds on this change until an addition is invalidated due to area and power constraints, the architecture converges to the theoretical best possible latency, or no operator is left with a conflict that causes the time to cross the ALAP start time.
The MCR heuristics are shown in Algorithm~\ref{alg:fcf}.

\input{body/algorithms/firstconflict}

The rationale behind this heuristic is two fold.
First, if an operator's start time is beyond its ALAP schedule time, it would undoubtedly increase the overall latency of the graph execution.
Second, the forward and backward pass operators are arranged in a mirror dataflow; hence, resolving earlier conflicts in the forward pass can potentially resolve conflicts in the backward pass, significantly improving the throughput of the overall training iteration. 

\niparagraph{Greedy Scheduler for Heuristics}:
The heuristics employs a greedy scheduler through the algorithm.
Operators are scheduled greedily, meaning they are scheduled if all their predecessors are completed and the required core is available. 
If two operators are ready but insufficient cores are available, the order is determined based on operator criticality.
The combination of ASAP/ALAP schedules defines the slack for each operator's start time.
Operators with zero slack are the most critical.
For the remaining operators, higher slack means lower priority, and vice versa. To reduce idle time, a low-priority operator can be added before a critical operator when it doesn't impact the critical operator's start time.
As we traverse the graph, the order of operators within a core/unit adhere to the dependencies in the graph.
All the operators within a single core/unit are executed in-order.
Dependencies across units are maintained using a semaphore block.

%% file: body/algorithms/firstconflict.tex
\RestyleAlgo{ruled}

 \DontPrintSemicolon

\begin{algorithm}
\footnotesize

\newcommand{\funccommd}[1]{{\footnotesize\textcolor{blue}{#1}}}
\newcommand{\mycommfont}[1]{{\scriptsize\itshape\textcolor{brown}{#1}}}

\caption{Mirror Conflict Resolution Heuristics}\label{alg:fcf}

\KwIn{$G(V,E)$ \quad \quad \quad \quad  \quad \quad \quad \quad \mycommfont{// Annotated Operator graph}}


\KwIn{\emph{TC-Dim, VC-Width}  \quad \quad \quad \mycommfont{// Current config of the architecture}}

\KwIn{$Constraints$ \quad \quad \quad \quad \quad \quad \mycommfont{// Area and power constraints}}

\KwResult{$\#TC, \#VC$}

$config_{curr}$ = \emph{$<$1, TC-Dim, 1, VC-Width$>$} \quad \mycommfont{// Initialized with 1 core}

$schedule_{time} = \funccommd{GreedyScheduler}(G,\;config_{curr})$

\While{$config_{curr} \neq config_{prev}$}
{
    \ForEach{$node \in G(V,E) $}
    {
        $delay_{critical} = schedule_{time} - ALAP_{time}$\;
        \If{$\funccommd{CheckResourceConflict}(node) \& \; delay_{critical}>0$}
        {
            break
        }
    }

    $config_{prev} = config_{curr}$\;
    \mycommfont{// Adding the necessary core and updating the configuration}
    $config_{curr} = \funccommd{AddCoreCheckConstraints}(node, constraints)$

    $schedule_{time} = \funccommd{GreedyScheduler}(G,\;config_{curr})$\;
    \If{$\funccommd{CheckRuntimeIsWorse}(schedule_{time})$}
    {
        return $config_{prev}$
    }
}

\end{algorithm}

%% file: body/ilp.tex
\subsection{Integer Linear Programming Formulation}

The heuristics search for the number of cores takes a deliberate approach towards prioritizing resources towards critical operators.
To offer formal guarantees of optimality as an alternative to heuristics, we also formulate our search of the number of cores as an ILP.
ILP similar to the heuristics is bounded by critical-path's best latency as that is the limitation of the model. 
Even with this bound, the integer program is co-optimizing the \# of cores and the schedule of the operators, thus can take a non-trivial amount of time.

\niparagraph{Problem Definition:}
Let $ G(V,E) $ denote the operator graph with vertex set $ V $ and edge set $E$.
A $v$ denotes a single vertex in $V$ and $e$ for a single directed edge in $E$.
$ \Delta v $ denotes the estimated latency of each operator $v$.
Possible types of cores is denoted by $ C $, and in this work we assume $ C = [Tensor\;Core,\;Vector\;Core] $. 
However, our ILP formulation works for any set $C$.
For a core $c \in C$, the variable $ x(c) $ denotes the number of cores of type $ c $ our solution uses, and we assume that $ x(c) \geq 1 $ by preprocessing the input.
The function $ M: V \rightarrow C $ gives a mapping of operators $ V $ to computational core $ C $; an operator $v \in V$ needs to be processed on the core $M(v)$.
Let  $ A(c) , P(c) $ denote the area utilization and power consumption of each unit of core $c$, and let $A , P$ denote the total area and power constraints. We require that the total area and power used by all computational cores is at most $A , P$. 
The main decision variables are $y_{(v,t)} $, that indicate when the operator $v$ is scheduled. We assume that time is slotted and entire DAG can be feasibly scheduled in $T$ time slots. We get an estimate of $T$ by doing a binary search.  For an operator $v$, $ y_{(v,t)} = 1 $ only if $v$ starts its execution at time slot $t$. If $ y_{(v,t)} = 1 $, then it means that operator $v$ is scheduled on core $M(v)$ in the contiguous set of time slots between $[t, t + \Delta v - 1]$.

\niparagraph{ILP Objectives:}
As we aim to minimize the training time, area, and power, we formulate a multiple objective ILP.

\begin{itemize}
    \item First objective minimizes the training iteration time by tuning the number of cores. We formulate it as follows:
    
    \vspace{-1ex}
    \begin{mini}
          {}{\sum_{t \in T}{} t \cdot y_{(v^*,t)}}{}{}{}
    \end{mini}
    \vspace{-2ex}
    
\end{itemize}

\begin{itemize}
    \item Second objective minimizes the area and power consumption whilst keeping it within the constraints. 
    
    \vspace{-1ex}
    \begin{mini}
	  {}{f(z) = \sum_{c}{} x(c) \cdot A(c), \; f(p) = \sum_{c}{} x(c) \cdot P(c)}{}{}
	\vspace{-0.5ex}  \addConstraint{f(z) \leq A, \; f(p) \leq P}{}{}
    \end{mini}
    \vspace{-3ex}
\end{itemize}

\niparagraph{ILP Constraints:}
The constraints ensure a valid schedule of operators is obtained that respects the graph dependencies.

\begin{itemize}
    \item First set of constraints enforce that each operator gets scheduled only once and is executed non-preemptively.
    \vspace{-0.5ex}
    \begin{equation}
        \vspace{-1ex}
        \sum_{t \in T}{} y_{(v,t)} = 1 \quad \quad \forall \; v \in V
    \end{equation}
    
    \item Next we enforce capacity constraints. We ensure that the total number of operators that require computational core $ c $ at any time $ t $ is $\leq$ the tuned number of computational cores.
    \vspace{-0.5ex}
    \begin{equation}
        \vspace{-1ex}
        \sum_{v \in V, M(v) = c}{} \sum_{t' = t - (\Delta v - 1)}^{t} y_{(v,t')} \leq x(c) \quad \forall \; t \in T, \; c \in C
    \end{equation}
    
    Above constraint implies that if an operator $v$ has a start time $t'$ (that is, $y_{(v,t')} = 1$) then it would require core $M(v)$ for the entire duration of $[t', t' + \Delta v -1]$.
    \item Finally, we want the operators to be scheduled in order of their precedence within the operator graph.
    \vspace{-0.5ex}
    \begin{equation}
            \vspace{-1ex}
        \sum_{t \in T}{} t \cdot y_{(v',t)} - \sum_{t \in T}t \cdot y_{(v,t)} \geq \Delta v \quad \forall \; e:(v,v') \in E,  v \rightarrow v'
    \end{equation}

\end{itemize}

\niparagraph{ILP Outputs:}
As output, the ILP provides the optimal number of cores and optimal schedule (variable $ x(c) $) required for the workload within the area and power constraints. 
We obtain the optimal schedule from variable $ y_{(v,t)} $ of each operator $ v $.

%% file: body/algorithms/config_gen.tex
\RestyleAlgo{ruled}

\begin{algorithm}
\caption{Configuration Pruner for Tensor Core}\label{alg:config_gen}

\footnotesize

\newcommand{\funccommd}[1]{{\footnotesize\textcolor{blue}{#1}}}
\newcommand{\mycommfont}[1]{{\scriptsize\itshape\textcolor{brown}{#1}}}

\KwIn{$MaxTCDim$ \quad \quad \quad \mycommfont{// TC dimension range}}
\KwIn{$VCWidth$ \quad \quad \quad \quad \mycommfont{// Vector core width for this pruning}}
\KwIn{$StepSize$ \quad \quad \quad \quad \quad \mycommfont{// Step size to decrement the dimensions}}
\KwIn{$HysLevels$ \quad \quad \quad \quad  \mycommfont{// Hysteresis level as input for the pruning}}

\DontPrintSemicolon

$TCDimsToExplore.\textcolor{violet}{append}(MaxTCDim)$ \quad \mycommfont{// Starting Config}

$MinRuntime = \funccommd{CriticalPathSearch}(CurrentTCDim, VCWidth)$\;

\While{$TCDimsToExplore \neq  \emptyset$}
{
    $CurrentTCDim = TCDimsToExplore.\textcolor{violet}{pop}()$
    
    \mycommfont{// Generates configs for the next level and ignores duplicates.}
    
    $NewTCDims = \funccommd{GenerateNextDim}(CurrentTCDim, \; StepSize)$\; 
    
    $Runtimes = \funccommd{CriticalPathSearch}(NewTCDims,\;VCWidth)$;
    
    \If{$min(Runtimes) < MinRuntime$}
    {
        $MinRuntime = min(Runtimes)$\;
        $TCDimsToExplore.\textcolor{violet}{append}(\funccommd{GetBetterConfigs}(NewTCDims))$
    }
    \ElseIf{$\funccommd{Check}(HysLevels)$}
    {
        $TCDimsToExplore.\textcolor{violet}{append}(NewTCDims)$
    }
}
\end{algorithm}

%% file: body/tables/search_space.tex
\begin{scriptsize}
\begin{table}[!ht]
\centering
\caption{Search space comparisons, excluding per-operator mapping.}
\resizebox{1\columnwidth}{!}{
\begin{tabular}{l c c c c c}
 \hline
 \multirow{2}{*}{\textbf{Model}} & \multirow{2}{*}{\textbf{Exhaustive}} & \multicolumn{2}{c}{\textbf{ILP}} & \multicolumn{2}{c}{\textbf{Heuristics}} \\

 &  & \textbf{Unpruned} & \textbf{Pruned} & \textbf{Unpruned} & \textbf{Pruned} \\
 \hline
 MobileNet\_v3 & $10^{38}$ & $10^{24}$ & $10^{14}$ & $10^{21}$ & $10^{10}$\\
 Inception\_v3  & $10^{39}$ & $10^{25}$ & $10^{14}$ & $10^{22}$ & $10^{12}$\\
 ResNeXt-101  & $10^{40}$ & $10^{26}$ & $10^{15}$ & $10^{23}$ & $10^{13}$\\
 BERT-Large & $10^{40}$ & $10^{26}$ & $10^{16}$ & $10^{23}$ & $10^{13}$\\
 \hline
\end{tabular}}
\label{tab:search_space}
\end{table}
\end{scriptsize}

\vspace{-1ex}

%% file: body/distributedsearch.tex
\section{Global Search for Accelerators}
\label{sec:global}

The Global Architecture Search module performs an optimization to determine the architecture for a set or single workload across the stages of distributed training.

\niparagraph{Partitioning the model.}
Operator graph is partitioned using existing device placement techniques~\cite{pip, piper, flexflow}. 
This work specifically handles pipeline and model parallel split which impact the operator graph that executes on each device. Data parallel is a replicated pipeline and hosts the same graph across.
As a proof of concept, \wham includes a memory-balanced splitter that partitions the graph based on HBM capacity and memory requirements of training. 
Based on user inputs such as pipeline training scheme (e.g., Gpipe or Pipedream), pipeline depth, and batch size, along with model properties such as parameter and activation size, the splitter determines the memory footprint of training and partitions the model.
For model parallel, the per stage operator graphs are based on Megatron style splits per device. 
The tensor model parallel width is given as an input to the \wham search.

\niparagraph{Networking.}
Pipeline parallel training only requires activations to be transferred from one device to the next, and \wham accounts for the latency of data transfers across neighboring accelerators via interconnects defined in the system configuration.
Model parallel requires collective operators such as allreduce in forward and backward pass to collect the intermediate results.
Similar to prior works~\cite{flexflow, piper}, we assume a homogeneous network where all devices communicate with each other. 
Hierarchical or multi-level network topologies are not considered within the scope of this work.

\subsection{Configuration Search using Top-k Designs}

This search obtains \emph{top-k} designs for each partition of the operator graph per device using the search described in Section~\ref{sec:local}. 
Selecting $k$ instead of a 1 design per stage as the top design for a particular stage in the pipeline may not necessarily yield a balanced pipeline, which is crucial for achieving high throughput and utilization of the entire system. 
%
%
The search for architectures for distributed training presents a challenge in that each model has multiple designs to select from, resulting in $k \times s$ architectures for an s pipeline depth execution. 
When optimizing for a set of workloads, the number of architectures grows to $k \times s \times m$, where m is the number of models. Evaluating every possible configuration and selecting the best across models would be time-consuming. 
To address this challenge, the global module employs a top-level pruning policy similar to Configuration Pruner in Section~\ref{sec:pruner}.

\begin{figure}
	\centering
	\includegraphics[width=0.45\textwidth]{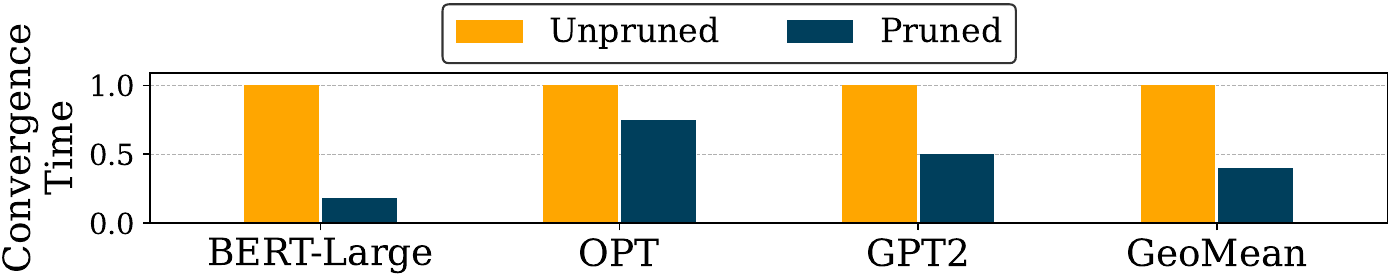}
	\caption{Convergence time comparison to search for a global design in pipeline parallel training with a pipeline depth of $32$ and $k=10$.}
	\label{fig:dist_prune}
\end{figure}

This top-level pruner takes unique configurations from the $k \times s \times m$ designs to construct a search tree.
Each configuration comprises both the dimensionality and number of cores. 
Each level in the tree contains designs of the same area, with root node as the the smallest design. 
Evaluating smaller to larger architectures ensures that larger configurations in the lower levels of the tree that consume more energy but do not offer better performance can be pruned.
The pruner eliminates sub-trees if a larger configuration is worse in metric across all the models.
Alternatively, if all children are worse than the direct parent, subtrees are pruned once the evaluation reaches the hysteresis level and all evaluated child configurations are worse.
As illustrated in Figure~\ref{fig:dist_prune}, the pruned distributed search converges \convgtimeimprdist faster than the unpruned search, which evaluates all top-k configurations across all the models.

%% file: body/evaluation.tex
\section{Evaluation}
\label{sec:evaluation}

\subsection{Experimental Setup}

\niparagraph{Models:}
\wham is evaluated across a diverse set of workloads, such as Vision~\cite{inception, mobilenetv3, resnet, resnext, vgg}, Translation~\cite{gnmt}, and Language Modeling~\cite{bert, opt, gpt}.
Table~\ref{tab:models} shows the details of models and their training configurations.
Additionally, we evaluated the performance of \wham for distributed training of language models.

\niparagraph{Software Implementations:}
We use PyTorch-1.9~\cite{pytorch} to obtain the operator graphs~\cite{torchviz}.
We use readily available training scripts such as torch-vision for image classification, GNMT~\cite{gnmt} from NVIDIA~\cite{nvidia}, and language models from huggingface~\cite{huggingface}.
\wham is executed on an 8-core Intel Xeon E5-2673 CPU with Haswell architecture and 28 GB DDR4 main memory.
The ILP is solved using Gurobi~\cite{gurobi}).

\niparagraph{Performance Metric:}
\wham optimizes for relevant training metrics, such as throughput or energy efficiency.
For efficiency, as established by prior works, we use the correlated Perf/TDP for efficiency~\cite{tpuv4i, fast} due to the proprietary nature of TCO.
With throughput as the metric, \wham designs maximize end-to-end throughput within power and area constraints. 
With Perf/TDP, \wham designs maximize Perf/TDP while maintaining a user-specified minimum end-to-end throughput.

\subsection{Baselines}

We compare \wham against two types of baselines: prior search frameworks ConfuciuX~\cite{confuciux} and Spotlight~\cite{spotlight} and established hardware architectures for deep learning.
This allows the evaluation to establish both the efficacy of the search technique and the generated accelerators.
All baselines assume an HBM of 16~GB~\cite{tpuv2} and a bandwidth of 900~GB/s~\cite{tpuv3}.

\input{body/tables/models_new}

\niparagraph{Prior Frameworks.}
\wham is the first framework to explore the design space for training. 
To compare with other approaches, we extend two state-of-the-art frameworks, ConfuciuX and Spotlight, to incorporate training.
While ConfuciuX and Spotlight perform search over forward pass (inference) for GEMM and Convolution operators, using reinforcement learning and Bayesian optimization, respectively, we extend these frameworks, \textbf{ConfuciuX+} and \textbf{Spotlight+}, to support backward pass and weight update pass for all GEMM and Convolution operators.
%
ConfuciuX+ selects the largest configuration across forward, backward, and weight update passes similar to its original version. 
In contrast, Spotlight+ optimizes for architecture for the backward pass and weight update pass, in addition to the forward pass.
To consider the point-wise vector operations ignored by both frameworks, we use the same vector core width as suggested by the framework for the tensor core.

\niparagraph{Comparison against hand-optimized accelerators:}
We assess \wham against hand-designed accelerators, specifically TPUv2-like~\cite{tpuv2} and NVDLA-like~\cite{nvdla} designs, including their corresponding dataflows. 
We use a scaled-up version of NVDLA to incorporate training.
This design has one $256 \times 256$ tensor core and one 256 wide vector core (\emph{$<$1, 256 $\times$ 256, 1, 256$>$}).
TPUv2~\cite{tpuv2} contains 2 computational units, each having tensor core with $128 \times 128$ systolic array and 128 wide vector core (\emph{$<$2, 128 $\times$ 128, 2, 128$>$}).

\niparagraph{Compiler and runtime optimizations.}
Both \wham and baselines use common compilation and runtime techniques in deep learning. 
Op-fusion is applied when a convolution or GEMM operator is followed by an activation function~\cite{tvm, polymath} to reduce data movement across the memory subsystem. 
Additionally, runtime data reuse allows in-flight and ready-to-schedule operators to share intermediate results, reducing costly round trips to HBM as data is directly consumed on the chip.

\subsection{Results for Individual Accelerator Search}

\wham can configure for either \emph{throughput} or \emph{Perf/TDP}. 
For each metric, it can search for a configuration specific to a single workload, \wham-individual, or a common configuration that works for a set of workloads, \wham-common. 
Larger workloads OPT, GPT2-XL and GPT3, are only evaluated for distributed training. 

\begin{figure}
	\centering
	\includegraphics[width=0.48\textwidth]{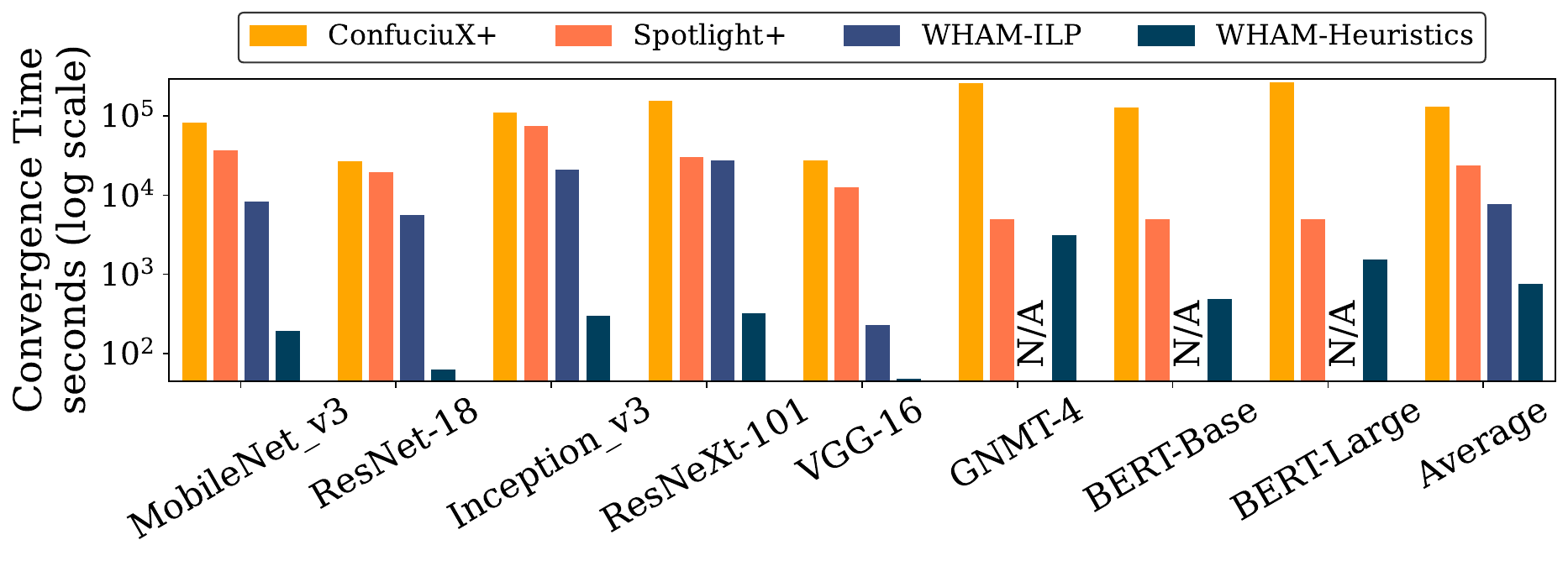}
	\caption{Convergence time comparison of \wham with prior frameworks. N/A does not converge in 7 days.}
	\label{fig:solving_time}
\end{figure}

\begin{figure*}[t]
	\centering
	\includegraphics[width=0.85\textwidth]{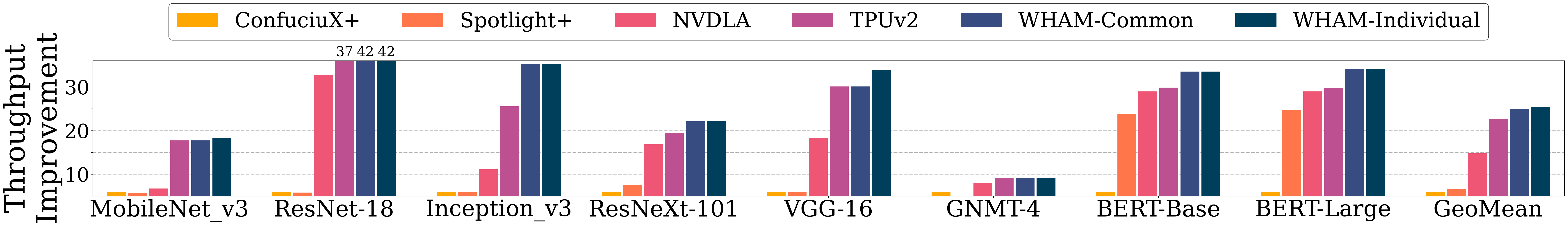}

    \caption{Comparison of \wham generated acclerators \emph{optimized for throughput} against hand-optimized accelerators and framework suggested designs. All results are compared to ConfuciuX+ generated design.}
    \label{fig:perf_comp}
\end{figure*}

\begin{figure}
	\centering
	\includegraphics[width=0.48\textwidth]{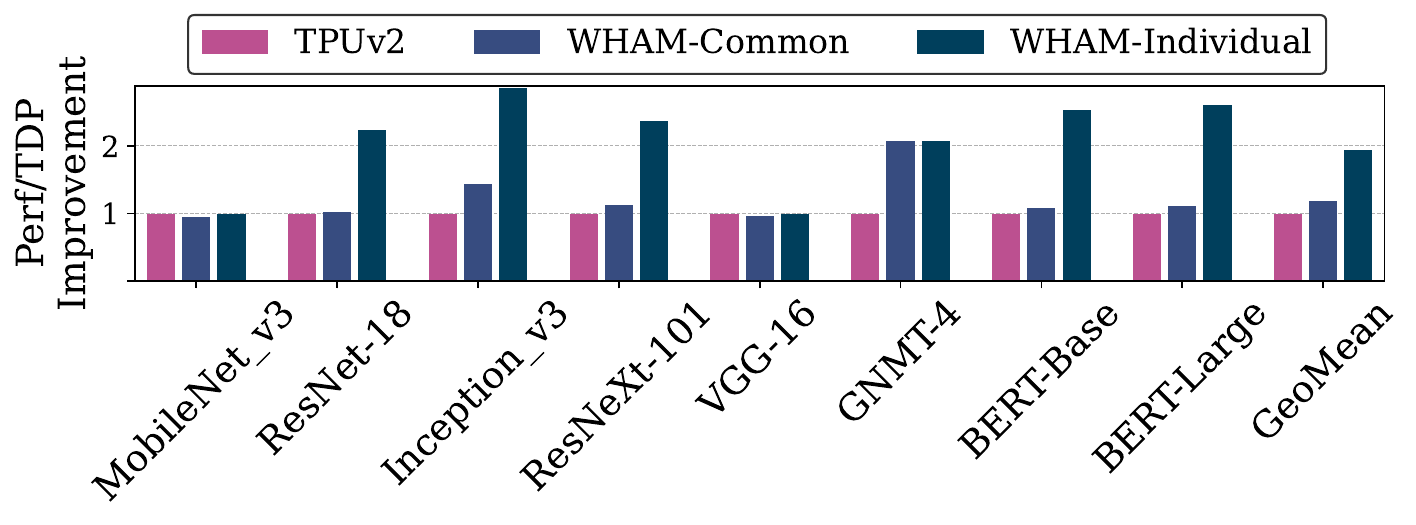}
    \vspace{-4ex}
	\caption{Comparison of \wham generated designs \emph{optimized for Perf/TDP} with TPUv2 as baseline.}
	\label{fig:perf_tdp}
\end{figure}

\niparagraph{Convergence Time.}
Figure~\ref{fig:solving_time} compares the convergence time of \wham ILP and heuristics against prior frameworks, ConfuciuX+ and Spotlight+.
We run \wham and prior frameworks for 500 iterations and compare their wall clock times.
On average, \wham converges \convgtimeimprconfuciux and \convgtimeimprspotlight faster than ConfuciuX+ and Spotlight+, respectively.
This is because \wham employs a novel algorithmic technique that deliberately reduces the search space using the pruner and the critical-path-based approach. In contrast, prior approaches use reinforcement learning, genetic algorithm, and Bayesian optimization techniques and scale the problem proportional to the problem size.

For ConfuciuX+, the RL converges to a local minima relatively quickly, while the genetic algorithm takes a long time to fine-tune the minima. 
Spotlight+ reduces the search space by removing duplicate problem dimensions in a DNN graph and thus converges faster, especially for language models with replicated transformer layers.
However, Spotlight+ does not prune the architectural search space like \wham. 
We observed that the ILP in \wham could not converge within seven days for a single iteration of architectural configuration for language and translation models due to the large size of the DNN graph.

\input{./body/tables/single_acc}

\niparagraph{Architecture comparison with Throughput as Metric.}
Table~\ref{tab:local_opt} shows the architectures proposed by each framework and \wham.
Tensor Core L1-reg has a size of 512 Bytes while vector core and tensor core L2-SRAM sizes are shown in the table.
Figure~\ref{fig:perf_comp} presents the throughput improvement of \wham-  individual and -common over prior frameworks and hand-optimized accelerators with \emph{throughput} as the optimization metric.
\wham-individual is compared against ConfuciuX+ and Spotlight+ generated architectures and TPUv2 and NVDLA to \wham-common.

On average, \wham-individual provides \avgthroughputindividualconfuciux and \avgthroughputindividualspotlight throughput improvement over ConfuciuX+ and Spotlight+, respectively.
ConfuciuX+ and Spotlight+ generated configurations are inefficient mainly due to the large training design space.
Due to their search techniques, they fail to converge on a design suitable across forward, backward and parameter update pass and mostly rely on the biggest configuration to accommodate the training complexities.

The \wham-common design addresses the needs of all the evaluated workloads, and offers \avgthroughputcommonnvdla and \avgthroughputcommonTPU higher throughput over NVDLA and TPUv2, respectively.
The reason for these benefits is the improved utilization of the cores and the exploited concurrency across operators, allowing them to be scheduled in parallel over multiple cores.
For example, in the BERT model, the QKV projection in each encoder layer can be executed in parallel across three tensor cores, which justifies the architectural configuration for this model.
For BERT-Base and BERT-Large models, such parallelism is the source of up to 53\% of performance improvement over the TPUv2 baseline.
For workloads without any branching structure (MobileNet\_v3, VGG-16, etc.), the main source of performance improvement is better utilization of the core.
\wham-individual, however, can offer \avgthroughputindividualnvdla and \avgthroughputindividualTPU higher benefits in comparison to NVDLA and TPUv2, respectively.
These configurations are specialized for a single model and employ model-specific spatial unrolling of the output and input feature interactions across tensor core dimensions.

\niparagraph{Architecture comparison with Perf/TDP as Metric.}
Figure~\ref{fig:perf_tdp} shows the Perf/TDP benefits of \wham's proposed architecture compared to TPUv2 like design.
\wham optimizes for Perf/TDP with a throughput constraint of TPUv2. 
While designs generated by ConfuciuX+, Spotlight+, and NVDLA are not compared in the Figure due to their focus on latency-bound inference, it is worth noting that both \wham-common and \wham-individual provide orders of magnitude higher Perf/TDP than all these designs.
This is because these frameworks heavily optimize for tensor-core only operators and often select the largest design, leading to low utilization but high energy.
In contrast, \wham deliberately optimizes for training and considers a metric of interest.
Compared to TPUv2, \wham-common provides \avgperftdpcommonTPU better Perf/TDP than TPUv2, because two cores do not limit it and can exploit operator concurrency beyond two for many models.
For certain models, \wham-individual does not offer any higher benefits as models exhibit little to no branching.

\subsection{Global Search for Distributed Training}

This section compares the architectures generated across LLMs for pipeline and model parallel execution.

\niparagraph{Pipeline parallel training.}
We compare global search results across various generated designs: \wham-common, a common architecture across pipeline stages addressing all models, \wham-individual is tailored to each model but homogeneous across its pipeline, and \wham-mosaic top-1 design for each stage in the pipeline for every model, resulting in a heterogeneous pipeline. 
Our results are presented using a pipeline depth of 32, GPipe~\cite{gpipe} pipeline strategy, and activation stashing.

\begin{figure}
	\centering
	\includegraphics[width=0.48\textwidth]{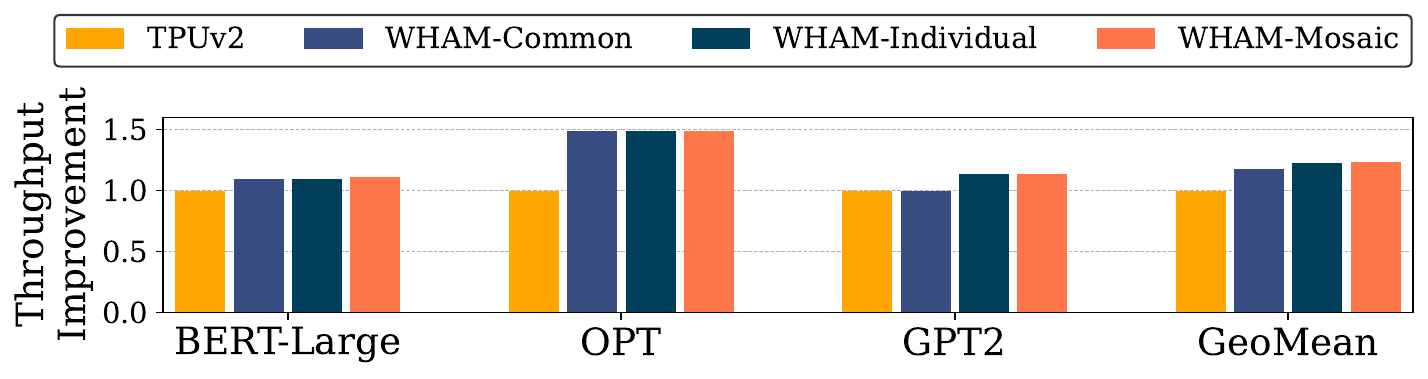}
	\caption{Throughput comparison of \wham designs for pipeline parallel training compared to TPUv2, optimized for throughput.}
	\label{fig:dist_throughput}
\end{figure}

\niparagraph{Architecture comparison with Throughput as a Metric.}
Figure~\ref{fig:dist_throughput} compares the training throughput of the \wham optimized accelerator with the TPUv2 accelerator, the best-performing baseline.
On average, we observe a throughput improvement of \avgdistthroughputcommon, \avgdistthroughputindividual, and \avgdistthroughputmosaic for the Common, Individual, and Mosaic configurations, respectively, compared to TPUv2.
Among the generated designs, \wham-individual provides the most significant benefit because it is specific to each model.
Language models have repeated transformer layers, resulting in similar dataflow properties across model partitions. 
As a result, \wham-individual can capture a common design across those stages to cater to those properties. 
This is also why \wham-mosaic's heterogeneity only provides modest benefits over \wham-individual.

\begin{figure}
	\centering
	\includegraphics[width=0.48\textwidth]{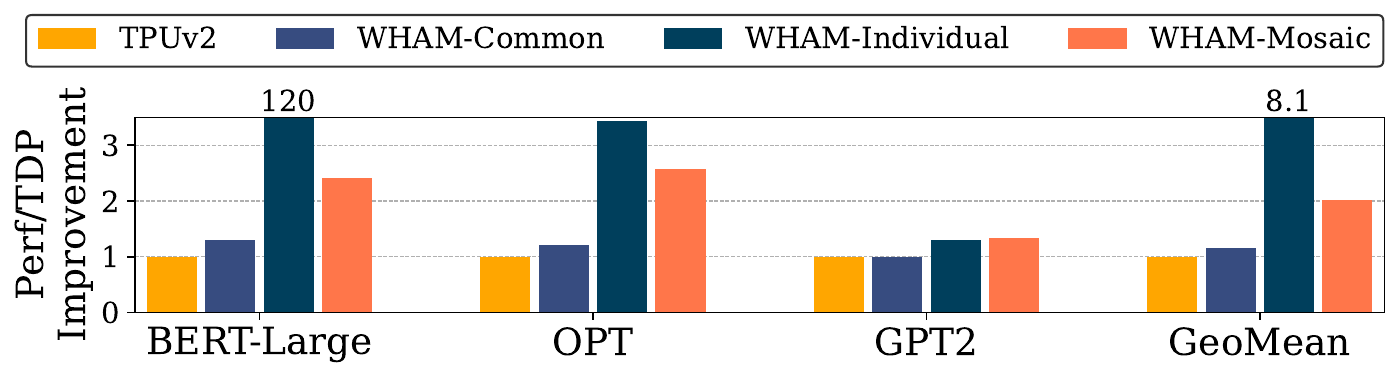}
	\caption{Perf/TDP comparison for pipeline parallel training with \wham designs compared to TPUv2 and optimized for Perf/TDP. Normalized to TPUv2 TMP-1, PP-64.}
	\label{fig:dist_perf_tdp}
\end{figure}

\niparagraph{Architecture comparison with Perf/TDP as a Metric.}
Figure~\ref{fig:dist_perf_tdp} compares Perf/TDP achieved by \wham generated designs when optimized for this metric and the minimum throughput of the TPUv2 like architecture.
On average, the configurations generated by \wham exhibit a Perf/TDP improvement of 1.6$\times$, \avgdistperftdpindividual, and \avgdistperftdpmosaic for the Common, Individual, and Mosaic configurations, respectively, compared to TPUv2 design. 
It is important to note that when optimizing for Perf/TDP, the top-1 architecture optimized for each pipeline stage, \wham-Mosaic, may not yield a better end-to-end metric. 
This is because each pipeline stage chooses the best architecture for its own stage, but due to the bottleneck stage, it may not contribute to higher throughput while consuming more energy due to the larger area.
In contrast, \wham-individual considers all pipeline stages to accommodate the end-to-end metric and can generate a homogeneous architecture that provides better Perf/TDP. 
Its worth noting that \wham-common must be generalized across workloads.

\begin{figure}
	\centering
	\includegraphics[width=0.48\textwidth]{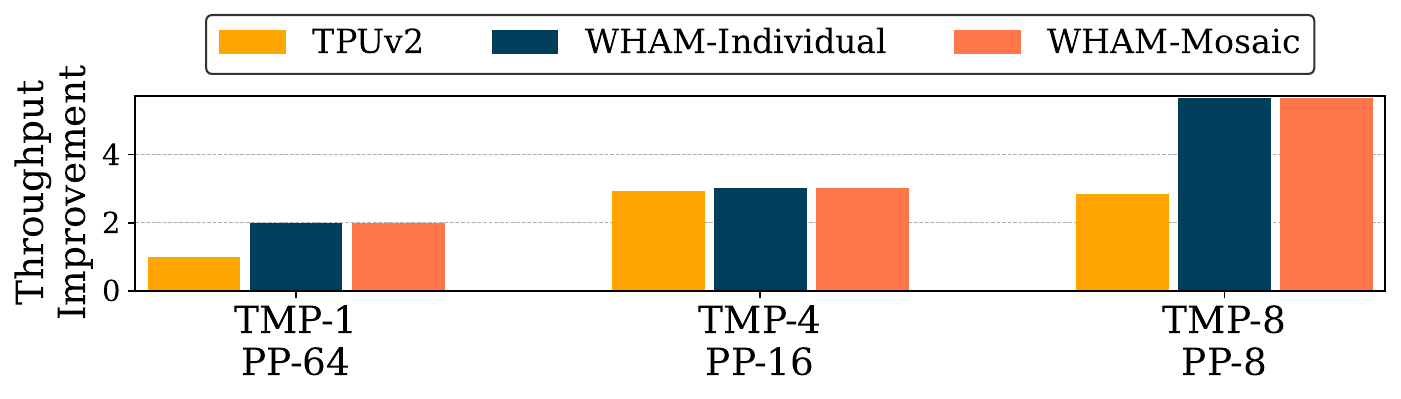}
    \vspace{-5ex}
	\caption{GPT3 throughput comparison between various tensor model (TMP) and pipeline parallel configurations, when using \wham designs, in contrast to TPUv2. The total devices is 64.}
	\label{fig:tmpc}
	\vspace{-3ex}
\end{figure}

\niparagraph{Model and Pipeline Parallel Training.}
To address training of large models, model parallelism (TMP) is used with pipeline parallelism.
Although our evaluation explores architectures for Megatron-style split, we can support any TMP or pipeline parallel strategy by obtaining the corresponding operator graph that resides on each device.
Figure~\ref{fig:tmpc} illustrates the throughput improvements achieved by \wham configurations compared to TPUv2 like design, as TMP scales from 1 to 8.
The total number of devices involved in training is 64.
\wham proposed architecture provides \tmpcimprv throughput improvement over TPUv2 architecture with TMP and pipeline parallelism of 8. 
As all the stages in GPT3 are uniform due to the model structure, \wham individual and mosiac results are identical.

\begin{figure}[h!]
	\centering
	\includegraphics[width=0.48\textwidth]{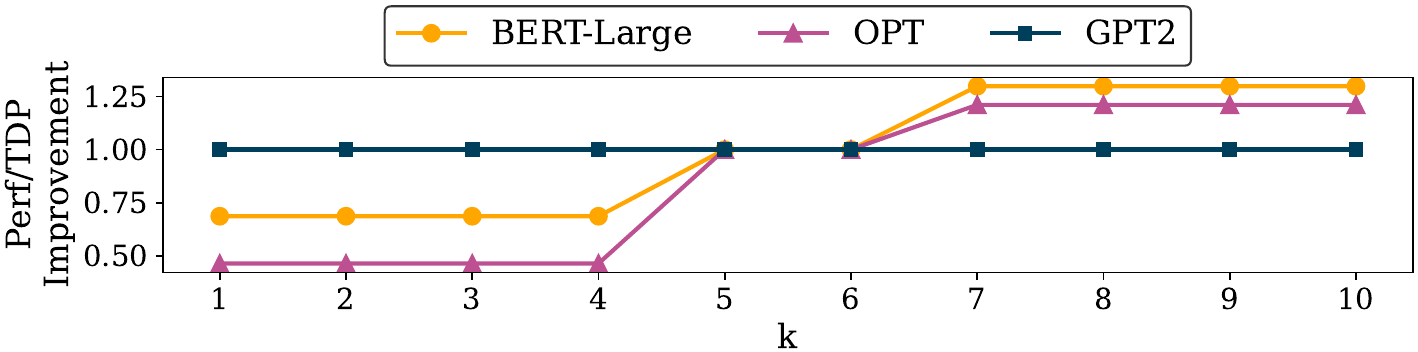}
    \vspace{-5ex}
	\caption{Top-k hyper-parameter sweep, and its impact on Perf/TDP for \wham-Common (all results compared to TPUv2).}
	\label{fig:top_k}
    \vspace{-3ex}
\end{figure}

\niparagraph{Top-k hyper-parameter search.}
We sweep the top-k hyper-parameter generated for each pipeline stage across three LLMs to determine the optimal value for distributed pipeline parallel training.
As Figure~\ref{fig:top_k} shows, naively selecting the top-1 design does not always yield the best metric, however, 
we observe diminishing returns as the Perf/TDP improvements saturate after $k$ = 10.

%% file: body/tables/models_new.tex
\begin{scriptsize}

\newcommand\ExtraSep
{\dimexpr\cmidrulewidth+\aboverulesep+\belowrulesep\relax}

\begin{table}
\centering
\caption{DNN models and their training configurations.}
\resizebox{1\columnwidth}{!}{
\begin{tabular}{l l c c c}
\hline
\multirow{2}{*}{\textbf{~Task}} & \multirow{2}{*}{\textbf{Model}} & \textbf{Model} & \textbf{Hyper} & \textbf{Number of}
\\
 &  & \textbf{Parameters} & \textbf{Parameters} & \textbf{Accelerators}
\\
\hline
\multirow{5}{*}{\begin{tabular}{l} Image \\[\ExtraSep] Classification \end{tabular}} & MobileNet\_v3~\cite{mobilenet} & 24~M & batch size:~128 & 1 \\
\cline{2-5} & ResNet-18~\cite{resnet} & 30~M & batch size:~128 & 1 \\
\cline{2-5} & Inception\_v3~\cite{inception} & 43~M & batch size:~64 & 1 \\
\cline{2-5} & ResNeXt-101~\cite{resnext} & 87~M & batch size:~16 & 1 \\
\cline{2-5} & VGG-16~\cite{vgg} & 141~M & batch size:~64 & 1 \\

\hline
\multirow{2}{*}{~Translation} & \multirow{2}{*}{GNMT-4~\cite{gnmt}} & \multirow{2}{*}{70~M} & batch size:128 & \multirow{2}{*}{1} \\
 & & & hidden size:~512 & \\
\hline
\multirow{16}{*}{\begin{tabular}{r} Language \\[\ExtraSep] Modeling \end{tabular}} & \multirow{2}{*}{BERT-Base~\cite{bert}} & \multirow{2}{*}{110~M} & batch size:~4 & \multirow{2}{*}{1} \\
& & & sequence length: ~512 & \\
\cline{2-5} & \multirow{4}{*}{BERT-Large~\cite{bert}} & \multirow{4}{*}{340~M} & batch size:~8 & \multirow{2}{*}{1} \\
 & & & sequence length:~128 & \\
\cline{4-5}
 & & & batch size:~1/32 & \multirow{2}{*}{1/32} \\
 & & & sequence length:~512 & \\
 \cline{2-5} & \multirow{3}{*}{OPT~\cite{opt}} & \multirow{3}{*}{1.3~B} & batch size:~ 32 & \multirow{3}{*}{32} \\
  & & & num layers:~24 & \\
  & & & attention heads:~32 & \\
\cline{2-5} & \multirow{3}{*}{GPT2 (XL)~\cite{gpt}} & \multirow{3}{*}{1.5~B} & batch size:~32 & \multirow{3}{*}{32} \\
 & & & sequence length:~512 & \\
 & & & attention modules:~48 & \\
 \cline{2-5} & \multirow{4}{*}{GPT3~\cite{gpt3}} & \multirow{3}{*}{175~B} & batch size:~4 & \multirow{3}{*}{64} \\
 & & & sequence length:~2048 & \\
 & & & num layers:~96 & \\
 & & & attention heads:~96 & \\
\hline
\end{tabular}}
\label{tab:models}
\end{table}
\end{scriptsize}

%% file: body/tables/single_acc.tex
\begin{scriptsize}
\newcommand\ExtraSep
{\dimexpr\cmidrulewidth+\aboverulesep+\belowrulesep\relax}
\begin{table}
\centering
\caption{Per accelerator architecture comparison. \wham architectures with Heuristics are \emph{optimized for throughput}. Designs are represented as \emph{$<$ \# TC, TC-DIM, \# VC, VC-Width $>$}.}
\resizebox{1\columnwidth}{!}
{
\renewcommand{\arraystretch}{1.2}
\begin{tabular}{ l | c | c | c | c | c}
 \hline
 \multirow{2}{*}{\textbf{Model}} & \multirow{2}{*}{\textbf{ConfuciuX+}} & \multirow{2}{*}{\textbf{Spotlight+}} & \multicolumn{3}{c}{\textbf{WHAM}}  \\
  \cline{4-6} & & & \textbf{L2 SRAM} & \textbf{Individual} & \textbf{Common}\\
 \hline
 
 MobileNet\_v3 & \multirow{8}{*}{\rotatebox[origin=c]{90}{\begin{tabular}{c}$< 1, 256 \times 256, 1, 256 >$\\[\ExtraSep] L2 SRAM : 32~MB \end{tabular}}} & $<1, 12 \times 512, 1, 12>$ & 8~MB & $<1, 256 \times 128, 1, 256>$ & \multirow{8}{*}{\rotatebox[origin=c]{90}{\begin{tabular}{c} $< 3, 128 \times 128, 3, 128 >$\\[\ExtraSep] L2 SRAM : 16~MB \end{tabular}}} 
 \\
 ResNet-18 &  & $< 1, 256 \times 240, 1, 256 >$ & 18~MB & $< 2, 128 \times 64, 2, 128 >$ & 
 \\
 Inception\_v3 & & $< 1, 128 \times 446, 1, 128 >$ & 8~MB & $< 4, 128 \times 64, 4, 128 >$ & 
 \\
 ResNeXt-101 & & $< 1, 244 \times 256, 1, 244 >$ & 6~MB & $< 2, 128 \times 64, 2, 128 >$ & 
 \\
 VGG-16 &  & $< 1, 128 \times 264, 1, 128 >$ & 32~MB & $< 1, 256 \times 128, 1, 256 >$ &
 \\
 \cline{1-1} \cline{3-5} 
 GNMT-4 & & $< 1, 60 \times 896, 1, 60 >$ & 8~MB & $< 3, 128 \times 64, 3, 128 >$ &
 \\
 \cline{1-1} \cline{3-5}
 BERT-Base & & $< 1, 64 \times 552, 1, 64 >$ & 8~MB & $< 3, 128 \times 64, 3, 128 >$ & 
 \\
 BERT-Large & & $< 1, 64 \times 960, 1, 64 >$ & 8~MB & $< 3, 128 \times 64, 3, 128 >$ & 
 \\
 \hline
 \end{tabular}
}

\label{tab:local_opt}
\end{table}
\end{scriptsize}

%% file: body/related_work.tex
\section{Related Work}
\label{sec:related}

\subsection{Architectural Search Frameworks}

Comprehensive architectural search techniques for deep learning are currently only focused on inference~\cite{fast, confuciux, prime, hasco, spotlight}.
FAST~\cite{fast} maps operators to Tensor and Vector cores for inference accelerators, using a black box optimizer~\cite{vizier} that generates the search hyperparameters.
It aims to utilize the extra global buffer memory for subsequent operators through fusion. 
However, in training, intermediate activations must be stashed for the backward pass, thus the extra memory might not be available for this optimization.
PRIME~\cite{prime} is an offline approach devised for inference that utilizes logged simulation data, to architect hardware accelerators without requiring new simulations. 
PRIME creates a cost function through a surrogate model but does not address how to generalize a single cost function for the forward, backward, and parameter update passes required in training.
Spotlight~\cite{spotlight} searches the HW/SW co-design space by injecting hand-provided domain information formulated as a Bayesian optimization.
Spotlight optimizes for layer-wise tensor core cost estimation targeting inference. 
ConfuciuX~\cite{confuciux} targets only inference and employs reinforcement learning and genetic algorithms to determine the number of PEs and local buffer size without considering vector operations.
It optimizes per layer and selects the largest design across layers for end-to-end inference execution.

Other works in this area~\cite{hasco, mbo, dnnweaver:micro, tabla:hpca, interstellar_dataflow_overrated} provide HW/SW solutions for dense tensor computation. 
Apollo~\cite{apollo} uses transfer learning, FlexiBO~\cite{flexibo_bayesian} and HASCO~\cite{hasco} apply Bayesian optimization, HyperMapper~\cite{hypermapper-random_forest} employs random forests, and others utilize genetic algorithms~\cite{cnn_design_genetic} for design space exploration. 
dMazeRunner~\cite{dmazerunner} and ZigZag~\cite{zigzag} focus mainly on large software design spaces. 
Design Space Exploration for recommendation models training~\cite{fae, hotline, adrec, slipstream} involves embedding exploration while federated learning~\cite{fluid} also benefits from DSE on each device. 

In contrast, \wham prioritizes training, which necessitates hardware accelerator optimized across forward, backward, and weight update passes. 
Furthermore, none of the mentioned works address distributed execution, whereas \wham performs architecture search for  pipeline parallel training.

\subsection{Mapping Frameworks}

Various mapping search frameworks~\cite{timeloop, marvel, mindmapping, gamma} determine data movement and compute placement across a design for a fixed architecture.
Marvel~\cite{marvel} optimizes dataflow for an architecture by reducing off-chip movements.
Timeloop~\cite{timeloop} employs random pruning to find the mapping for single operations (GEMM or CONV). 
MindMapping~\cite{mindmapping} is a gradient-based search method for mapping search exploration.
GAMMA~\cite{gamma} uses a genetic algorithm to develop an optimized mapping for a given layer.

In contrast, \wham aims to identify the architecture design while considering dataflow. 
To optimize the architecture for training, \wham utilizes existing open-source dataflow mapping search techniques for deep learning operations.
It can integrate with any dataflow search framework for its architecture search.

%% file: body/conclusion.tex
\section{Conclusion}

\wham is the first work to perform architecture search for hardware accelerators in a pipeline and model parallel setting.
This is an important problem as models are becoming larger and require multiple accelerators. 
\wham solves this problem via a multi-step approach.
The distributed search performs multiple isolated searches for an accelerator executing the model partition.
Each individual accelerator, the search takes a critical-path based algorithmic approach to determine the number of cores and buffers for each type.  
\wham then obtains the top-k designs for each accelerator and combines the results to determine the architectural configuration for a distributed setting.
This enables \wham to scale for training and search for accelerators across across a wide range of DNN workloads.